\documentclass[useAMS,usenatbib,usegraphicx]{mn2e}
\newcommand\degree{$^{o}$}
\newcommand\be{\begin{equation}}
\newcommand\ee{\end{equation}}

\newcommand\bea{\begin{eqnarray}}
\newcommand\eea{\end{eqnarray}}
\newcommand\HII{\hbox{H\,{\sc ii}}}
\newcommand\CII{\hbox{C\,{\sc ii}}}
\newcommand\CIIl{\hbox{[C\,{\sc ii}]}}
\newcommand\CI{\hbox{[C\,{\sc i}]}}

\newcommand\HI{\hbox{H\,{\sc i}}}

\newcommand\CO{$^{12}$CO}

\newcommand{\twCO}{$^{12}$CO}
\newcommand{\thCO}{$^{13}$CO}
\newcommand{\ext}{\mbox{T$_{\rm ex}$}}

\newcommand{\kms}{\mbox{km~s$^{-1}$}}
\def\cplus{C$^{+}$}
\newcommand{\cmcub}{\mbox{cm$^{-3}$}}
\newcommand{\cmsq}{\mbox{cm$^{-2}$}}
\newcommand{\Cone}{\mbox{[C\,{\sc i}]~$^3$P$_1$--$^3$P$_0$}}

\newcommand\araa{ARA\&A}%
\newcommand\apj{ApJ}%
\newcommand\apjl{ApJ}%
\newcommand\aap{A\&A}%
\newcommand\aaps{A\&AS}%

\title[ \CI\ observations toward Cas A]
{ C\,{\sc i} 492 GHz mapping toward Cas A}

\author[B. Mookerjea, N. G. Kantharia, D. Anish Roshi \& M. Masur] 
{B. Mookerjea$^{1,2}$\thanks{E-mail:bhaswati@astro.umd.astro.edu}, 
N. G.  Kantharia$^{3}$\thanks{E-mail:ngk@ncra.tifr.res.in} 
D. Anish Roshi$^{4}$\thanks{E-mail:anish@rri.res.in}, 
M. Masur$^{2}$\thanks{E-mail:masur@ph1.uni-koeln.de}\\
$^{1}$Department of Astronomy, University of Maryland, College
Park, MD 20742, USA\\
$^{2}$KOSMA, I. Physikalisches Institut, Universit\"at zu K\"oln,
Z\"ulpicher Strasse 77, 50937 K\"oln, Germany\\
$^{3}$National Centre for Radio Astrophysics, TIFR, Post Bag 3, Ganeshkhind, Pune 411007, India\\
$^{4}$Raman Research Institute, Sadashivanagar, Bangalore 560
080, India \\}

\begin{document}

\date{Accepted . Received ; in original form }

\pagerange{\pageref{firstpage}--\pageref{lastpage}} \pubyear{}

\maketitle

\label{firstpage}

\begin{abstract}

We have mapped the \Cone\ emission at 492~GHz toward the supernova
remnant Cas~A. We detect \CI\ emission from the periphery of the
diffuse Photon Dominated Region (PDR) covering the disk of Cas~A, as
traced by the carbon recombination lines, as well as from the denser
PDRs associated with the molecular clouds towards the south-east.  \CI\
emission is detected from both the Perseus and Orion arm molecular
clouds, with the $-47$~\kms\ Perseus arm feature being strong enough
to be detected at all positions. We estimate the C/CO relative
abundance to be 0.2 at the position of the identified CO clouds and
$>1$ for most of the cloud. Here we show that the distribution of
\CI\ emitting regions compared to the \cplus\ region and molecular
cloud is consistent with a scenario involving PDRs. Using physical
models for PDRs we constrain the physical properties of the \CI\
line-forming regions. We estimate  the densities of the \CI\ emitting
regions to be between 10$^2$ and 10$^3$~\cmcub. Based on rather high
volume filling factors ($\sim 50$\%) we conclude that \CI\ emission
mainly arises from diffuse neutral gas in the Perseus arm.

\end{abstract} 

\begin{keywords}
ISM:general -- ISM:lines and bands -- ISM:molecules -- ISM:clouds --
radio lines:ISM -- radio lines:general -- Galaxy:general. 
\end{keywords}

\section{Introduction}

Molecular clouds are embedded in the diffuse interstellar medium (ISM)
of the Galaxy where most of the gas is in an atomic form. The atomic
gas is mostly located in two phases that are in pressure equilibrium,
the cold neutral medium (CNM) with temperatures of $T\sim 100$~K and
the warm neutral medium (WNM) with $T\sim 10^4$~K
\citep{kulkarni1987,dickey1990,wolfire2003,mckee2004}. Some of the
cold atomic gas is also associated with the denser molecular clouds,
possibly forming an extended halo or envelope
\citep{andersson1993,moriarty1997}. 

Far-ultraviolet (FUV) photons ($6.0~<h\nu<~ 13.6$~eV) from OB stars
produce Photon Dominated Regions (PDRs) either at the interface
between the \HII\ region and the molecular cloud (classical high density
PDRs) or in neutral components (atomic or molecular) of the diffuse
interstellar medium (diffuse PDRs).  PDRs are defined as neutral
regions where the chemistry and heating are regulated by the FUV
photons  \citep{hollenbach1999}. Current understanding of PDRs suggest
chemical stratification in which with increasing depth from the
surface of the PDR, the dominant carbon-bearing species changes from
\cplus\ through C$^0$ to CO. Observations of \CIIl, \CI\
and CO lines tracing the different layers of PDRs is thus a useful
tool to constrain various conditions of the PDRs.


Ionized carbon in dense star forming regions is mainly traced
using the  $^2$P$_{3/2}$--$^2$P$_{1/2}$ fine structure line at
158~\micron\ \citep{howe1991,mookerjea2003}.  The radio observations
of carbon recombination lines (RRLs) provide a very good alternative
way to trace the spatial distribution of ionized carbon.  Classical
\CII\ regions which form part of high-excitation PDRs, are usually
identified through the observation of narrow (4 -- 10 \kms) carbon
RRLs at frequencies $>$ 1 GHz toward \HII\ regions
\citep{pankonin1977,wyrowski2000} and have been well studied.  The
diffuse \CII\ regions which form part of low-excitation PDRs, are
identified through  observations of carbon RRLs in absorption at
frequencies below $\sim$ 150 MHz and in emission above $\sim$ 200 MHz
\citep{payne1989}.  

Diffuse \CII\ regions were discovered more than two decades ago
through the detection of carbon RRL in absorption at 26.3 MHz
(C631$\alpha$) in the direction of Cas A \citep{konovalenko1980,
blake1980}. Since then several RRLs spanning over 14 to 1400 MHz have
been observed in this direction \citep[][and references
therein]{kantharia1998} and it remains the only diffuse \CII\ region
which has been studied and modelled in such detail using C RRL.  A
smooth transition of lines in absorption at frequencies below 115 MHz
to lines in emission at frequencies above 200 MHz has been detected
\citep[][see Fig.  1]{payne1989}.  Moreover, this is the only
direction in the Galaxy where high resolution 
($\sim$ 1\arcmin) image of the diffuse \CII\ region in carbon RRLs
near 332 MHz exist \citep[][see Fig 2]{kantharia1998}.
Modeling the line emission in this direction shows that the carbon
RRLs originate in cold regions with $T_e$ = 75 K and $n_e$ =
0.02~\cmcub\ \citep{kantharia1998}.   Several molecular line studies
have been performed towards the direction of Cas~A. These observations
do not detect much CO emission from the disk of Cas~A, but identify
molecular clumps located primarily to the south east of the disk
\citep{troland1985,wilson1993,liszt1999}.  The diffuse \CII\ regions
could  be coexistent with the cold neutral medium (CNM) that produce
\HI\ absorption toward Cas A \citep{payne1994} or with the molecular
component of the interstellar medium (ISM) \citep{ershov1987}.  In the
direction of Cas~A, a morphological comparison of C RRL distribution
with that of \HI\ and \CO\ indicate that the C RRL emission is more
likely associated with \HI\ gas \citep{kantharia1998}.

The upper state of the \Cone\ transition at 492~GHz is only 23~K above
ground. For this transition, the critical density for collisions with
H$_2$ molecules is only 1000~\cmcub\ \citep{schroeder1991}.  This
implies that \CI\ is easily excited and also the line is easily
detectable even when emitted by moderate density interstellar gas
exposed only to a radiation field equal to the mean interstellar
radiation field in the solar neighborhood. Thus \CI\ can be used as a
reliable tracer of the diffuse PDRs as well.  Here we present \CI\
mapping observations at 492~GHz in the direction of Cas~A in order to
probe whether the \CI\ emission like the C RRL emission arise solely
from the atomic CNM or the molecular phase also contributes to it. We
have compared the \CI\ observations with the C270$\alpha$, \HI, \twCO\
2--1 and \thCO\ 1--0 observations available in literature and have
used PDR models to explain the neutral carbon and CO emission wherever
the two spatially overlap.

\section{Observations and Data Analysis}
\label{sec:obs}

We have mapped the Cas A region in the fine structure transition
$^3$P$_1$--$^3$P$_0$ at 492~GHz of atomic carbon using the
SubMillimeter Array Receiver for Two frequencies (SMART; Graf et al.
2002) on KOSMA (Winnewisser et al.  1986), a 3-m submillimeter
telescope located on the Gornergrat in Switzerland. SMART is a
dual-frequency eight-pixel SIS-heterodyne receiver that observes
simultaneously at 4 positions (separated by 116\arcsec) on the sky at
two frequencies in the range 455-495~GHz and 795-882~GHz. The IF
signals are analyzed with array-acousto-optical spectrometers
(array-AOSs).  The array-AOS consists of 4 AOSs each with a bandwidth
of 1~GHz and a spectral resolution of 1.5~MHz (Horn et al.  1999).
This corresponds to a velocity resolution of 0.63~\kms\ at the
frequency of the \CI\ line.  Owing to technical difficulties the
higher frequency channel of SMART could not be used during our
observations. Typical receiver noise temperature achieved at the
center of the bandpass at 492~GHz is 150~K. Based on cross-scans on
Jupiter at 492~GHz we estimated the HPBW to be 55\arcsec\ and the beam
efficiency ($\eta_{\rm~mb}$) to be 50\%. 

We observed a fully sampled map centred at Cas A ($\alpha_{2000}$ =
$23^h23^m24^s$; $\delta_{2000}$ = $+58^{\circ}48'.9$), extending over
$\sim$ 6\arcmin $\times$ 7\arcmin. The observations were done in the
On-The-Fly (OTF) position switched mode, with the reference position
being 2\degree\ south of the map centre.  The map presented in
this paper required 5 complete coverages of the region and the total
integration time per position on the sky was 50 seconds. The
atmospheric calibration were done by measuring the atmospheric
emission at the reference position to derive the opacity (Hiyama 1998)
and the sideband imbalances were corrected for using standard
atmospheric models (Cernicharo 1985).  

Data was analyzed using the GILDAS \footnote{\tt
http://www.iram.fr/IRAMFR/GILDAS} spectroscopic data reduction
package.  Sinusoidal baselines were subtracted in order to correct for the
standing waves.

\section{Results}


\subsection{Velocity Structure in \CI}

\begin{figure}
\centering
\includegraphics[width=7.0cm,angle=0]{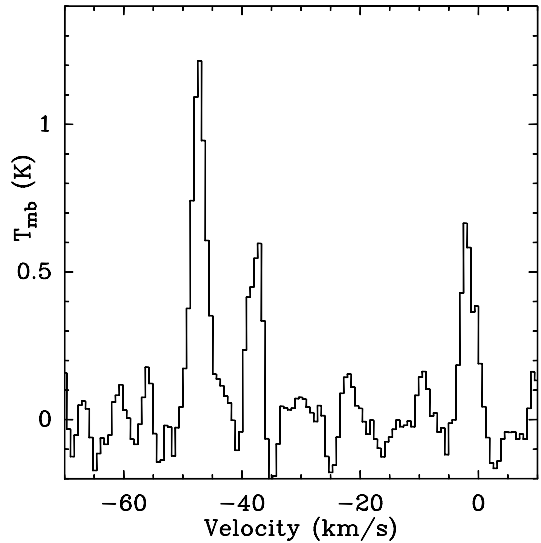}
\caption{\CI\ spectrum towards Cas~A averaged over part of the
observed region, limited in the west to $\Delta \alpha = -80$\arcsec\
and in the north to $\Delta \delta = 80$\arcsec. The Perseus arm and
Orion arm components are clearly detectable. The effective velocity
resolution of the spectrum is 0.63~\kms.
}
\label{avspec}
\end{figure}

Figure~\ref{avspec} shows the observed \CI\ spectrum averaged over
part of the map restricted in the west to $\Delta \alpha =
-80$\arcsec\ and in the north to $\Delta \delta = 80$\arcsec. We
identify mainly three strong emission features, at $-47$~\kms\ and
$-38$~\kms\ corresponding to the Perseus arm and at $-1$~\kms
corresponding to the local Orion arm. Since the $-38$~\kms\ feature
and the Orion arm feature do not appear at all positions within the
mapped region, the spectrum obtained by averaging over the entire
mapped region does not show them clearly. Thus, in order to provide a
fair representation of all velocity components detected in the region
we show here the average spectrum over the selected region.

\begin{table}
\centering
\caption{Results of  Gaussian fits to the average spectral
profiles of \CI\, \twCO\ 2--1, \thCO\ 1--0, C$270\alpha$ and
optical depth spectrum of \HI.}
\label{gaussfit}
\begin{tabular}{cccc}
\hline
Line & Component &  Centroid & Width    \\ 
     &           &   \kms & \kms    \\     \hline
\CI\ & Perseus arm &  $-47.4\pm0.1$ & $2.9\pm0.2$\\
     & Orion arm   &  $-1.7\pm0.1$ & $3.1\pm0.4$ \\
\hline
\twCO\ 2--1 &         & $-46.9\pm0.01$ & $3.0\pm0.05$\\
            & Perseus arm & $-40.3\pm0.01$ & $4.5\pm0.07$\\
            &         & $-36.2\pm0.03$ & $3.0\pm0.07$\\
            & Orion arm   & $-1.1\pm0.01$  & $2.1\pm0.02$ \\
\hline
\thCO\ 1--0 &  Perseus arm & $-46.8\pm0.05$ & $2.1\pm0.1$ \\
            &              & $-39.7\pm0.15$ & $6.1\pm0.3$ \\
            & Orion arm   & $-1.1\pm0.06$  & $1.8\pm0.14$ \\
\hline
C$270\alpha$ & Perseus arm & $-46.8\pm0.6$ & $5.1\pm0.9$\\
\hline
\end{tabular}
\end{table}

\begin{figure}
\centering
\includegraphics[width=7.0cm,angle=0]{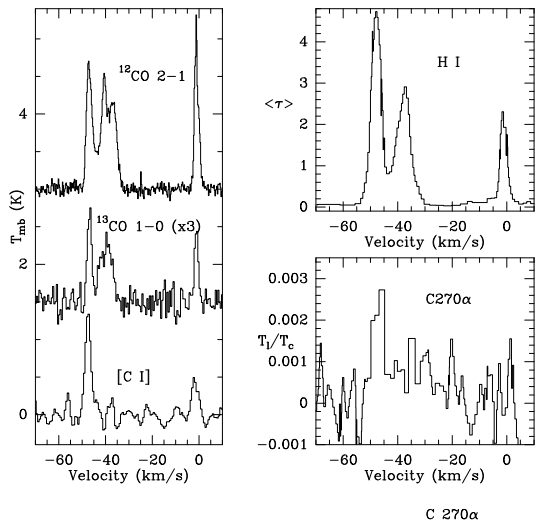}
\caption{Left panel:  Average spectra of \CI, \twCO\ 2--1, \thCO\
1--0 emission. Right panel: {\em Bottom:} Average spectrum of
C270$\alpha$ in units of line to continuum ratio  and {\em Top:}
Average spectrum of \HI\ optical depth.  All spectra are averaged over
the disk of Cas~A.
\label{cicospeccomp}}
\end{figure}

Figure~\ref{cicospeccomp} shows a comparison of the 
emission spectra of \CI, \thCO\ 1--0, \twCO\ 2--1, C270$\alpha$ (at
332~MHz) and the optical depth spectrum of \HI, all averaged over the
disk of Cas~A.  The CO spectra are from the observations by
\citet{liszt1999}, the C270$\alpha$ spectrum was observed by
\citet{kantharia1998}, and the \HI\ optical depth spectrum is from
\citet{schwarz1997}.  Table~\ref{gaussfit} presents the results of
fitting Gaussian components to the average emission spectra of the
different tracers.  The $-47$~\kms\ Perseus arm and the $-1$~\kms\
Orion arm features (not in C270$\alpha$) are detected most clearly in
all the emission and absorption spectra.  The \twCO\ 2--1 emission
shows two more Perseus arm features nominally around $-39$ and
$-36$~\kms.  The \CI, \thCO\ and \HI\ data do not resolve the two Perseus
arm features at $-39$ and $-36$~\kms, rather show a single feature at
$\sim -38$~\kms.  Although the C$270\alpha$ spectrum shows mainly the
$-47$~\kms\ feature, the same gas as traced by C RRLs with different
$\alpha$s also shows the $-38$~\kms\ feature both in emission and
absorption \citep[{\em cf.} Fig.~1,][]{payne1989}.



All tracers appear to show rather similar velocity components. Thus
based on velocity features it is not possible to isolate the
contribution of atomic and molecular components of the neutral ISM
towards the observed \CI\ emission.

Rest of the paper discusses the $-47$~\kms\ Perseus arm feature which
is detected in all tracers.

\subsection{\CI\ emission from the Perseus Arm at $-47$~\kms}

\begin{figure}
\centering
\includegraphics[width=8.0cm,angle=0]{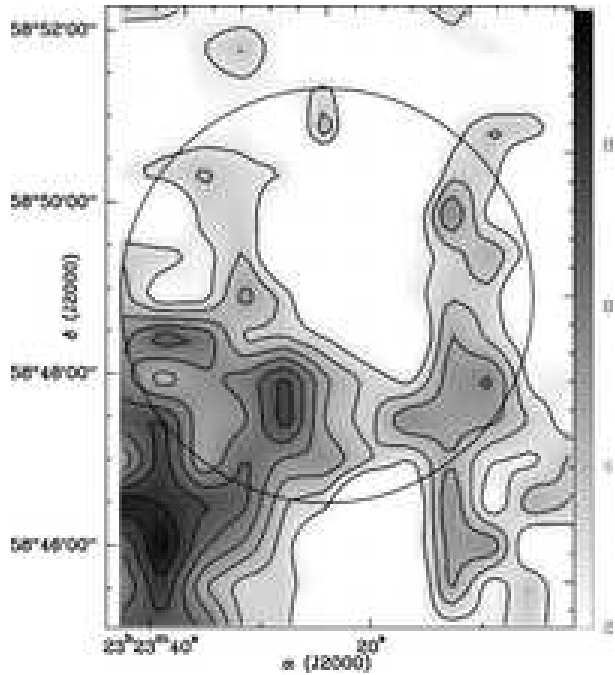}
\caption{Distribution of \CI\ emission from Cas~A. The intensities are
integrated between $-50.3$ and $-44.1$~\kms. Contours range 28 to 98\%
(in steps of 10\%) of the peak (9.7 K~\kms). Circle denotes the extent of the
Cas~A disk. }
\label{intmap}
\end{figure}

Figure~\ref{intmap} shows the intensity map of \CI\ emission from the
Perseus arm clouds along the line of sight to Cas A (the circle
denotes the extent of Cas~A), integrated between the velocities of
$-50.3$ and $-44.1$~\kms.  The observed \CI\ emission toward Cas~A
consists of two distinct morphological features. The first feature is
the bright emission detected towards the south-east and the second is
the filamentary structure to the west of the map extending almost
north-south.  There is almost no emission detected from the northern
part of the disk of Cas~A.

\begin{figure}
\centering
\includegraphics[width=8.5cm, height=4.5in]{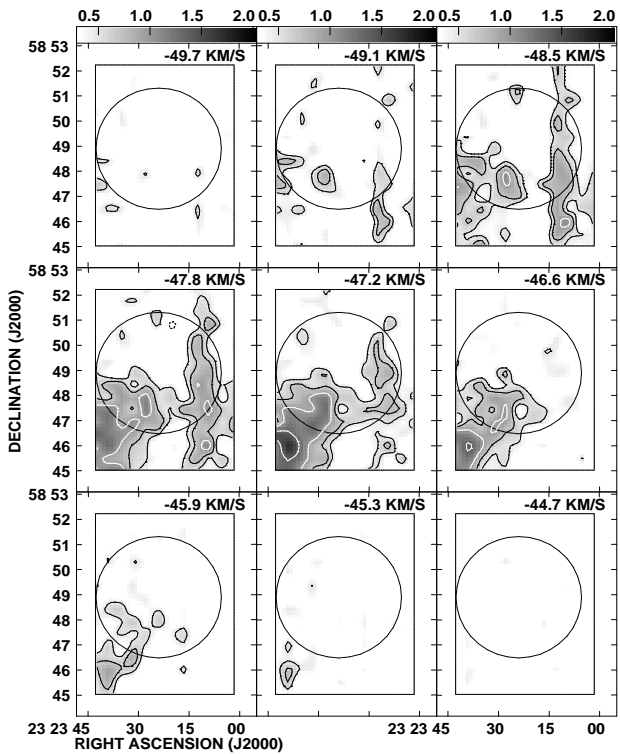}
\caption{Channel maps of CI emission.  The circle denotes the extent of Cas~A. 
Velocity is labelled in each panel. The grey scale ranges from 0.32 to 2 K and 
contours are plotted for 0.48, 0.68, 0.96 and 1.35 K. Note the wider south-east
component of \CI.}
\label{chanmap}
\end{figure}

Figure~\ref{chanmap} shows the channel maps of \CI\ emission between
velocities $-49.7$~\kms\ to $-44.7$~\kms.  It shows that the emission
from the cloud to the south-east covers the largest range of
velocities between $-48.5$ to $-45.9$ \kms. The linear
filamentary extension in the west in contrast is detected only between
$-49.1$ to $-47.2$ \kms.  

The morphology of \CI\ emission rules out interaction of this gas with
the SNR.  As shown in Fig.~\ref{chanmap}, the \CI\ emission is seen
mostly from the southern part of the supernova remnant; the eastern
part coincides with the molecular cloud seen in that region
\citep{wilson1993} which is also believed to lie in the Perseus
arm but not interacting with Cas~A. If there was interaction between
\CI\ line emitting region and Cas~A then it would lead to one of the
following being true: a) if the line emission is affected by the
continuum from Cas~A then it should show good morphological correlation
with the supernova remnant (SNR). However this is not the case here. b)
peculiar spectral features from the regions close to Cas~A e.g. broad
lines or stronger lines.  No such trend is seen and this also helps rule
out any interaction between the two. Thus, we believe that the \CI\
gas is not interacting with the supernova remnant.

\subsection{Comparison of \CI, C$270\alpha$ \& \HI\ distribution: Atomic Gas}

\begin{figure}
\centering
\includegraphics[width=8.0cm,angle=0]{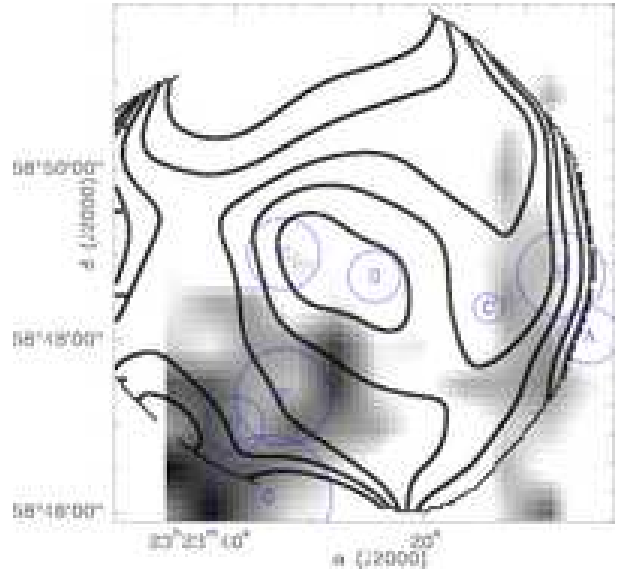}
\caption{Comparison of \CI\ (gray scale)  channel (of 1.4~\kms
width)  map with C$270\alpha$  optical depth (black contours) \citep{kantharia1998}
centred at a velocity of $-48.0$~\kms. The circles show the CO clouds 
identified by \citet{wilson1993}. The contour levels are 1, 2, 3, 4,
5, 7, 9 and 11 in units of 0.001. The greyscale ranges between 2 and
9.8~K~\kms.}
\label{ci_c270}
\end{figure}


Figure~\ref{ci_c270} shows the \CI\ emission superposed on the
contours of C$270\alpha$ optical depth \citep{kantharia1998} both at
velocities of $-48$~\kms.  While the \CI\ emission overlaps partially
with the C$270\alpha$ emission, it also appears to surround the
C$270\alpha$ emitting region and there are some significant
differences in morphology between the two. The C$270\alpha$ peak
optical depth is observed from the central part of Cas~A and is not
coincident with the \CI\ peak.  The \CI\ peak is located to the south
of the C$270\alpha$ peak which possibly indicates the chemical
stratification from \cplus\ through C$^0$ to CO.

The location of the peak of the C RRL emission at the centre of the
SNR in Cas~A is indeed intriguing.  Assuming that the observed line
width of recombination lines at 34.5 MHz was entirely due to radiation
broadening, \citet{kantharia1998} put a lower limit on the distance of
115 pc between the line forming region and Cas~A indirectly ruling out
the association of this gas with the SNR. On the other hand,
\citet{kassim1995} have reported flatter spectrum in the central
arcmin or so of Cas~A at frequencies below 330 MHz. The common
location of the recombination line gas and the absorbing gas is
intriguing and argues for a common origin.  However,
\citet{kassim1995} argue that the thermal absorbing gas is located
within Cas~A in which case a common origin of the gas is ruled out.
Moreover the carbon recombination line forming gas has a continuum
optical depth of 6.4$\times10^{-4}$ at 74~MHz which is much less than
the optical depth (1.3) of the gas that \citet{kassim1995} detect.  We
conclude that the positional coincidence of the carbon recombination
line forming gas and the flat spectrum region towards Cas~A does looks
fairly intriguing and demands further investigation.

The line of sight to Cas~A has provided a uniquely detailed set of
observations of RRLs from highly excited states of singly ionized
carbon
\citep[e.g.][]{konovalenko1980,ershov1987,payne1989,anantha1994,kantharia1998}.
Interpretation of these RRLs using models considering
dielectronic-like recombination suggest that with the exception of the
recombination line width, all of the Cas~A carbon recombination line
and $\lambda$21~cm \HI\ absorption line data can be attributed to a
region where the physical conditions are typical of the cold neutral
medium of the ISM \citep{payne1994}.


We also compared the \CI\ emission with the \HI\ distribution across
Cas~A \citep{schwarz1997}.  \HI\ is observed in absorption across the
entire Cas~A disk and the morphology varies with the radial velocity
with maximum optical depth being observed around $-48.2$ \kms.
The peak absorption at this velocity arises in the south-east region
which roughly coincides with the peaks seen in \CI\ at similar
velocity.  \HI\ absorption is also seen from the western half of Cas~A
with the peak shifted slightly towards the south.  A similar peak is
also observed in \CI\ emission.

The \CI\ distribution is thus partly similar to both \HI\ and
C270$\alpha$ morphology.  We compare the \CI\ emission with the
molecular component of the ISM traced by the CO emission in the next
subsection.

\subsection{Comparison of \CI\ and molecular line emission toward 
Cas~A}

\begin{figure*}
\centering
\includegraphics[width=17cm,angle=0]{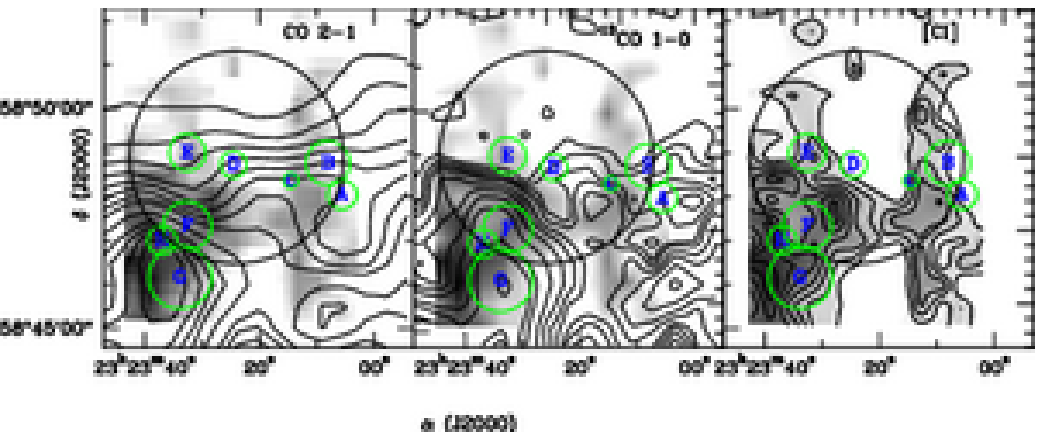}
\caption{Integrated intensity distribution of \CI\ (Grey) overlayed
with integrated intensities of \thCO\ 1--0 (left panel) and \twCO\
2--1 (middle panel) \citep{liszt1999}.  The \thCO\ 1--0 data has a
resolution of 60\arcsec. The \twCO\ 2--1 dataset has
been smoothed to a resolution of 60\arcsec\ for comparison. The 
large circle denotes the
angular extent of Cas~A. Extreme right panel shows both in grayscale
and  contours the integrated intensity of \CI. The smaller circles in
all panels show the position and
sizes of CO clouds identified by \citet{wilson1993}.
\label{cicocomp}}
\end{figure*}

The left and middle panels of Fig.~\ref{cicocomp} show the
distribution of \thCO\ 1--0 and \twCO\ 2--1 intensities (in contours)
integrated between velocities of $-50.3$ and $-44.1$~\kms\
\citep{liszt1999}, superposed with the \CI\ distribution (in grey).
While the original resolution of the \thCO\ 1--0 map is 60\arcsec, the
\twCO\ 2--1 data has been smoothed to 60\arcsec. In an earlier paper
\citet{wilson1993} observed the CO (and \thCO) emission from the
Perseus arm clouds in the direction of Cas~A, with an angular resolution
$\sim 22$\arcsec.  The CO clouds identified by \citet{wilson1993} are
also marked by circles in Fig.~\ref{cicocomp}.  We note that the CO
data from \citet{liszt1999} do not show the clumped structure observed
by \citet{wilson1993} and we attribute it to the difference in resolution
between the two datasets.


Figure~\ref{cicocomp} shows that the \CI\ emission is detected from
most of the CO emitting clouds and the \CI\ peaks lie at the outer
boundaries of the clumps.  Substantial \CI\ emission is detected with
no corresponding CO emission from the diffuse PDRs to the north. 
The extreme right panel of Fig.~\ref{cicocomp} shows the integrated
\CI\ intensities overplotted with the CO clouds identified by
\citet{wilson1993}. It clearly shows that although \CI\ emission is
detected from most of the clouds, the peaks of \CI\ emission tend to
lie at the boundaries of the CO clouds. We point out here that the CO
clouds detected by \citet{wilson1993} towards the western edge of the
map belong to the $-36$~\kms\ Perseus arm feature, in contrast to the
$-47$~\kms\ feature in \CI\ emission being discussed in this paper. 
The relative position of the \CI\ and CO emission peaks is consistent
with the clumpy PDR scenario, in which PDRs form on the surface of the
molecular clumps that are embedded in an interclump medium of lower
density in which CO is dissociated to atomic carbon.

From the above comparisons it is clear  that \CI\ traces both CO and
cold \HI\ morphologies partially and is consistent with the PDR
scenario when compared with the \cplus\ regions traced by the carbon
recombination lines (e.g. C$270\alpha$) and the molecular clouds
traced by CO.  We note here, that absorption studies of C~RRLs
necessarily imply that the emitting gas lies in front of the strong
continuum emitting SNR, no such restriction is however applicable to
either the CO or the \CI\ lines. Thus, beyond the similarities in the
emission velocities of \CI\, CO and the C$270\alpha$ and the
morphological similarity to a PDR-like structure, the present
observations cannot conclusively rule out the possibility that the
molecular (and neutral) and the ionized gas may not be occupyng the same 
volume of gas.

\section{Neutral carbon and CO column densities}
\label{sec_colden}

We have estimated the C$^0$ and CO column densities based on LTE
approximation. For these calculations we have used the observed \CI\
and \thCO\ 1--0 line intensities and an excitation temperature, \ext,
of  20~K.  Further we have assumed both \CI\ and \thCO\ 1--0 to be
optically thin.  The assumed \ext\ of 20~K is justified because based
on the \twCO\ and \thCO\ observations the kinetic temperatures of the
molecular clumps in the Perseus arm along the direction of Cas~A is
$\sim 20$~K \citep{wilson1993}.  


In the Perseus arm clouds with velocity $\sim -47$~\kms\ we estimate
the column density of neutral carbon, N(C), to be between 2$\times
10^{16}$~\cmsq\ and 1.3$\times 10^{17}$~\cmsq. Near the \CI\ peaks in
the south-east and also to the west, N(C) is larger than
7~10$^{16}$~\cmsq, while to the north and in the periphery of the \CI\
emitting shell the column densities are close to 3~10$^{16}$~\cmsq.

Within the region mapped by us in \CI, the estimated \thCO\ column
density (in a beam of 1\arcmin) varies between 5$\times 10^{14}$~\cmsq\ and
9$\times ~10^{15}$~\cmsq. This corresponds to N(CO) between
3.0$\times ~10^{16}$~\cmsq\ and 5$\times ~10^{17}$~\cmsq, assuming
[\twCO/\thCO]=60 . The column density of \thCO\ derived
from the 1\arcmin\ dataset of \citet{liszt1999} is slightly lower than the
values obtained by \citet{wilson1993}, but matches reasonably well
with the previous 1\farcm1 \thCO\ 1--0 observations by
\citet{troland1985}. 

Over the limited region in which the \thCO\ and \CI\ emissions
overlap, the C/CO abundance ratio varies between 0.08 and 1.8. At the
positions of the CO clouds F and G the C/CO abundance ratio is $\sim
0.2$, a value typically found at the centre of dense cloud cores
\citep{gerin1998,oka2004}.  In the periphery of the CO clouds the
value rises to 1.7 at positions where the \CI\ emission peaks : this
value is typical of the diffuse, translucent clouds
\citep{ingalls1997,bensch2003}. 


\section{PDR Interpretation of the observed \CI\ emission}
 
The observed morphology of \CI\ though not entirely correlated with
the emission from the atomic or the molecular phases of the ISM, is
consistent with the phenomenological understanding of PDR structure.
\CI\ is typically ubiquitous and towards the direction of Cas~A, it
appears to be arising from both the molecular clumps as well as the
diffuse PDR. Most importantly, the \CI\ emitting regions show no
evidence for interaction with the SNR in Cas~A. Here we explain the
observed \CI\ emission in terms of theoretical models for PDRs. We
have used the plane-parallel models by \citet{kaufman1999} for this
purpose.  The PDR model used here takes into account a detailed
chemical network, radiative transfer and thermal balance  and
considers a 1-dimensional slab of material exposed to a FUV radiation
field in a face-on geometry. The plane-parallel PDR model calculates
the intensity of various transitions of different chemical species as
functions of two parameters, the local hydrogen density ($n_{\rm
PDR}$) and the incident FUV radiation field. \citet{kaufman1999} have
created a database of PDR models, for local hydrogen densities $n_{\rm
PDR}$, between 10 \& 10$^7$~\cmcub\ and FUV radiation field between
0.3 to 3$\times 10^6$~G$_0$, where G$_0$ is the FUV flux measured in
units of the ``Habing field", which is taken to be
1.6$\times10^{-3}$~erg~cm$^{-2}$~s$^{-1}$~sr$^{-1}$ \citep{habing1968}.

In the absence of any nearby stars the far-ultraviolet (FUV) radiation
field illuminating the \CI\ emitting neutral atomic and/or molecular
phase of the ISM will be of the order of 1--5~G$_0$.  For the
following PDR analysis we assume the FUV flux to be $\sim 1$~G$_0$.

\subsection{Hydrogen densities}
\label{sec_pdr}

We have used the observed CI/\twCO 2--1 intensity ratios to derive an
independent estimate of the hydrogen volume densities in the \CI\
emitting regions using the PDR models.  For convenience from now on we
split our discussion into two parts, the first deals with the
south-eastern \CI\ emitting regions which are largely concomitant with
the CO emission and the second deals with the north-south stretching
filamentary emission to the west of the map. The aim of this
separation is also to look for possible differences between the two
parts. 

\begin{figure}
\centering
\includegraphics[width=7.0cm,angle=0]{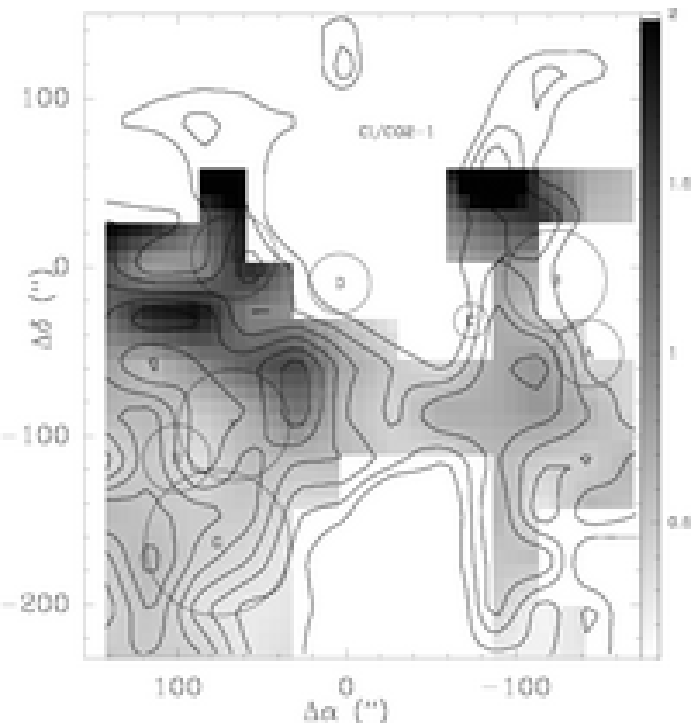}
\caption{\CI/\twCO 2--1 intensity ratio towards Cas~A. Only
positions with both \CI\ and \twCO\ 2--1 intensities larger than the
3$\sigma$ values  for the respective maps are included. The contours 
of \CI\ emission are overlayed for comparison. Also marked are the CO
clouds identified by \citet{wilson1993}. The map is centred at the nominal 
centre being used throughout the paper, viz., $\alpha_{2000}$ =
$23^h23^m24^s$; $\delta_{2000}$ = $+58^{\circ}48'.9$.
}
\label{cico21rat}
\end{figure}

Figure~\ref{cico21rat} shows the distribution of the CI/\twCO 2--1
intensity ratios in regions where both intensities are more than three
times the respective noise levels. Most of the pixels in both the
south-eastern cloud and the western filament show intensity ratios
between 0.5 and 1.0. At the position of the \CI\ emission peak to
the south-east the ratio is 1.3 and towards the northern edge of
the south-eastern cloud the ratio rises from 1.8 to 4. In the 
western filament also the value rises to as high as 4 towards
north. Clouds F, G and H show ratios of $\sim 0.5$, while cloud E
has a value of $\sim 1.3$. 

\begin{figure}
\centering
\includegraphics[width=7.0cm,angle=0]{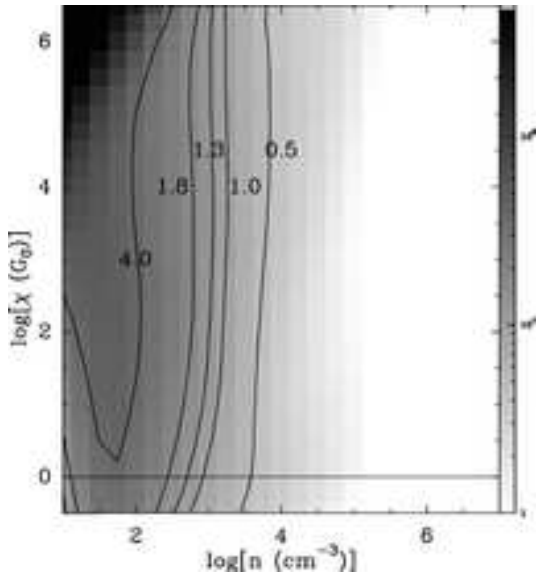}
\caption{PDR model calculations of variation of CI/\twCO 2--1
intensity ratios as a function of hydrogen densities and incident
UV radiation field \citep{kaufman1999}. The contours (marked with
their values) depict some of the typical ratios found toward Cas~A.
The horizontal line corresponds to the UV radiation field of
1~G$_0$.
}
\label{pdrdensity}
\end{figure}

Figure~\ref{pdrdensity} shows the variation of the CI/\twCO 2--1
intensity ratios as a function of the UV radiation field and the local
hydrogen densities as estimated by the PDR models by
\citet{kaufman1999}. For convenience we have drawn contours
corresponding to some of the typical ratios found in the direction of
Cas~A.  We note that according to the PDR models, the observed
CI/\twCO 2--1 intensity ratio constrains the local hydrogen density
almost independent of the UV radiation field assumed.  CI/\twCO 2--1
intensity ratio of 4 or more corresponds to densities less than
100~\cmcub. The most commonly seen ratio of 0.5 suggests densities of
the order of $10^{3.5}$~\cmcub. Thus the hydrogen density of the {\CI\
and CO emitting} gas varies from 100 to 3000~\cmcub.

Since the \CI\ emission only partially overlaps with the carbon
RRL emission as well as the CO emission in a manner
apparently consistent with the PDR scenario, it is interesting to
compare the neutral hydrogen densities in the different regions.
Based on latest models of carbon RRL emission, the
atomic hydrogen densities are estimated to be around 150~\cmcub, for
an electron temperature, T$_e$, of 75~K \citep{kantharia1998}. CO
observations suggest hydrogen volume densities of $\sim 10^3$~\cmcub\
with localized higher densities \citep{wilson1993}. We find that the
southern parts of the mapped region \CI\ emission arises primarily
from regions in which hydrogen is mostly molecular with average
densities of $10^{3}$~\cmcub. To the north, the densities are 
lower and the \CI\ emitting region has hydrogen primarily in the
atomic phase.  This is shown by the higher \CI/\twCO 2--1 intensity
ratios and the lower densities derived from the carbon RRL
emission which peaks in this region and is known to arise in the
atomic medium. The \CI\ emission thus not only morphologically but
also in terms of the continuity of density parameters acts as the
transition phase between the atomic and the molecular phase of the
neutral ISM.

\subsection{Volume and Area Filling factors of the  \CI\ emission}

\begin{table*}
\centering
\caption{Parameters observed and derived from PDR models at selected
positions in the Perseus arm. n$_{\rm PDR}$ is local hydrogen volume
density derived by comparing the observed CI\twCO 2--1 intensity ratios
with predictions of the PDR models. N(H$_2$) is the H$_2$ column density per
60\arcsec\ beam, $\phi_{\rm A}$ is the area filling factor of CI
emission defined as I$_{\rm CI,obs}$/I$_{\rm CI,th}$ and $\phi_{\rm V}$is the volume filling
factor defined as $n_{\rm PDR}$/$n_{\rm av}$.
Quantities with the subscripts {\em wilson} are from
\citet{wilson1993}.
\label{pdrtab}}
\begin{tabular}{lrr|lll|ll|ll|ll}
\hline
\hline
(1) &(2) & (3) & (4)& (5) & (6) &
(7) &(8) & (9) & (10)& (11) & (12)\\
Name & $\Delta \alpha$ & $\Delta \delta$ & I$_{\rm CI,obs}$ & 
N(H$_2$) &
n$_{\rm av}$ &
n$_{\rm PDR}$ 
& I$_{\rm CI,th}$ & $\phi_{\rm A}$ &  
$\phi_{\rm V}$ & N(H$_2$)$_{\rm wilson}$ & n(H$_2$)$_{\rm wilson}$\\
& \arcsec & \arcsec & K~\kms &10$^{20}$~\cmsq & \cmcub & \cmcub & K~\kms  & 
 & &10$^{20}$~\cmsq & \cmcub\\
\hline
Cloud E  &      58.&  0.  &  5.0 & 5.6  & 180   &300 & 10.9  &  
0.5  &     0.6 & 20 & $<$800\\
Cloud F   &     58.& -87.&   5.3& 39 & 1200 &2000 & 14.1 &  
0.4 &     0.6 & 40 & 1000 \\
Cloud G &       87.& -174.&  6.7 & 52 & 1600&5000 & 13.0  &  
0.5 &        0.3 & 90 & 2000 \\
\hline
CI Peak West &  -87. &  29.  &  2.9 & 5.6 & 180 &400  &  10.9 &  
0.3 &   0.5 &&\\
CI Peak SE   &  29. & -87. &   8.1 & 18 &560 &630 &  12.6  &  
0.6  &      0.9 &&\\
\hline
CI West Avg &  -87.& -58. &  4.8 & 11& 340 &  1778 & 14.1 &  
0.3 &     0.2 &&\\
\hline
\hline
\end{tabular}
\end{table*}

In order to quantify our results better we have selected a few
representative positions and derived densities at those positions for
an UV field of $\chi = 1$~G$_0$.  This analysis also aims to
characterize the clumpy nature of the emitting regions along the lines
of sight towards Cas~A. The positions chosen correspond to the centres
of the CO clouds E, F and G \citep{wilson1993} from which \CI\
emission was clearly detected. In addition, we have chosen the primary
\CI\ peak in the entire mapped region, lying in the south-eastern
molecular cloud and the secondary peak found to the north of the
filamentary emission to the west. Finally we have chosen a
representative position with average emission characteristics in the
western filament.  

Table~\ref{pdrtab} summarizes details of the selected positions and
several parameters observed and derived from PDR models and obtained
from literature.

As discussed in Sect.~\ref{sec_pdr} the local hydrogen volume
densities, $n_{\rm PDR}$, at the selected positions are estimated from
the observed CI/\twCO 2--1 intensity ratios using the PDR models (Col.
7, Table~\ref{pdrtab}). We find $n_{\rm PDR}$ for the CO clouds F and
G to be $\sim 10^3$~\cmcub, while cloud E has a much lower density of
only 300~\cmcub. These densities are in agreement with the values
derived by \citet{wilson1993} using much higher resolution (22\arcsec)
observations (Col. 12 Table~\ref{pdrtab}). At the position of the two
\CI\ peaks to the south-east and in the western filament the densities
are $\sim 500$~\cmcub, while for most of the positions on the western
filament the density is $\sim 10^3$~\cmcub. This implies that the high
\CI\ intensities arise from the more diffuse interclump medium.

The N(H$_2$) per 60\arcsec\ beam (Col. 5, Table~\ref{pdrtab}) for
the individual positions have been estimated as described in
Sec.~\ref{sec_colden}.  For the clouds E \& G the N(H$_2$) per beam
(60\arcsec) are lower by more than a factor of  $3$ than the N(H$_2$)
measured per 22\arcsec\ (Col. 11, Table~\ref{pdrtab}) by
\citet{wilson1993}, while for cloud F the two numbers agree. At those
positions where the \thCO\ 1--0 emission lies below the noise limits
of the dataset, we have assumed the noise level to derive
N(H$_2$)$<5.6\times 10^{20}$~\cmsq.  Using the hydrogen column
densities and assuming that the cloud extends the same distance along
the line of sight as the linear size of the beam we have derived the
beam-averaged volume densities ($n_{av}$; Col. 6,
Table~\ref{pdrtab}).  We note here that n(H$_2$)$_{\rm wilson}$
(Col. 12, Table~\ref{pdrtab}) has also been derived from the
N(H$_2$)$_{\rm wilson}$ in the same way.

The volume filling factors of the emitting clouds can be estimated as
$\phi_{\rm V}$=$n_{\rm PDR}$/$n_{\rm av}$.  For most of the positions
except for Cloud G and the average position in the western filament
the volume filling factors  are higher than 50\% (Col.~10,
Table~\ref{pdrtab}). This is in contrast to the typical volume filling
factors of 5\% as found in the Galactic star forming regions
\citep{kramer2004}. The observed high volume filling factors thus
indicate the rather diffuse nature of the \CI\ emitting regions.

Comparison of the absolute \CI\ intensities, I$_{\rm CI,th}$ (Column
7, Table~\ref{pdrtab}), estimated by the PDR models which best
reproduce the observed CI/\twCO 2--1 ratios, with the observed \CI\
intensities (I$_{\rm CI,obs}$; Col.~4, Table~\ref{pdrtab}) provides
information about the area filling factors (Col.~9,
Table~\ref{pdrtab}) of the \CI\ emitting clouds. The area filling
factor is defined as: $\phi_{\rm A}$=I$_{\rm CI,obs}$/I$_{\rm CI,th}$.
At the selected positions in the direction towards Cas~A the \CI\ area
filling factor is on an average equal to 0.5.

\section{Summary}

We have used the \CI\ emission at 492~GHz to probe the PDRs seen in
the direction of the supernova remnant Cas~A. We detect \CI\ emission
at velocities around $-47$ and at $-39$~\kms\ from the Perseus arm
clouds, and at $-1$~\kms\ from the local Orion arm cloud.  The
\CI\ emission does not show any strong morphological correlation with
the continuum emission from Cas~A.  Further, we do not detect any
obvious broadening or strengthening of spectral lines closer to the
continuum peak. Both of these conclusively rule out that the \CI\
emitting gas is interacting with the SNR in Cas~A.

We detect \CI\ emission from both the diffuse CNM (as traced by the C
RRLs) and the denser molecular cloud.  Morphological comparison of the
\CI\ emission with the carbon RRL (C270$\alpha$), \HI\ optical depth
and CO distribution appear to be consistent with a PDR scenario in
which the dominant carbon-bearing species changes from \cplus\ in the
diffuse region through C$^0$ to CO in the more dense molecular clumps.
We note with caution that despite the morphological and kinematic
evidences the present observations cannot rule out the possibility
that the \CI\ and C~RRL emitting regions may not completely occupy the
same volume of space.

We estimate the C$^0$ column density, N(C), to be between 2~10$^{16}$
and $1.3~10^{17}$~\cmsq. At the position of the previously identified
CO clouds we estimate the  C/CO abundance ratio to be around 0.2,
similar to the typical value found in dense cloud cores. Most of the
\CI\ emitting region however shows C/CO abundance ratios $>1$, typical
of diffuse, translucent clouds.

We have derived hydrogen volume densities of the \CI\ emitting regions
from the \CI/\twCO 2--1 intensity ratios using PDR models. We find
that the local hydrogen densities vary from $\sim 10^2$~\cmcub\ near
the \CI\ peaks to $10^{3.5}$~\cmcub\ around CO peaks. This also shows
a continuity in the density structure in the region between the the
diffuse \CII\ region and the denser molecular clouds. We estimate the
area filling factor of the \CI\ emitting regions to be  $\sim
0.3$, while the volume filling factor is on an average $>30$\%. This
further suggests that most of the \CI\ emission stems from the diffuse
PDR also traced by the carbon recombination lines.

\section*{Acknowledgments}
We thank the referee, Jay Lockman, for comments which improved the
clarity and presentation of the paper.
We thank H. Liszt for allowing us to use the CO datasets.
This material is based upon work supported by the Deutsche
Forschungs Gemeinschaft (DFG) via grant SFB494 and the
National Science Foundation under Grant No. AST-0228974.  This
research has made use of NASA's Astrophysics Data System.

\label{lastpage}

\end{document}